\documentclass[prd,twocolumn,superscriptaddress,altaffilletter,amssymb,showpacs,nofootinbib]{revtex4}
\usepackage[dvips]{graphicx}
\usepackage{amsmath}
\usepackage[latin1]{inputenc}
\newcommand{\be}{\begin{equation}}
\newcommand{\ee}{\end{equation}}
\newcommand{\bea}{\begin{eqnarray}}
\newcommand{\eea}{\end{eqnarray}}

\PassOptionsToPackage{linktocpage}{hyperref}
\usepackage[draft=false]{hyperref}
\usepackage{threeparttable}
\usepackage[draft=false]{hyperref}
\usepackage{threeparttable}

\begin{document}

\title{Higgs inflation on the brane}

\author{Dagoberto Escobar}\email{dagoberto.escobar@reduc.edu.cu} 
\affiliation{Departamento de F\'isica, Universidad de Camagüey, A.P. 74650, Cuba.}

\date{\today}

\begin{abstract}
We analyze the slow-roll inflation mechanism in brane framework with a real Higgs field confined on the brane.  We prove that inflation occur for field value below the 4-dimensional Planck scale and  produce cosmological perturbations in accordance with observations.  Through the amplitude of the scalar perturbation produced during inflation we could determine the self-coupling constant of the Higgs field which is not predicted by the fundamental theories. Also it was found the solutions to the motion equations on the brane.
\\

KeyWord: Higgs inflation, Higgs particle, cosmological perturbation, Randall-Sundrum braneworld.
\end{abstract}

\pacs{04.20.-q, 04.20.Cv, 04.20.Jb, 04.50.Kd, 11.25.-w, 11.25.Wx, 95.36.+x, 98.80.-k, 98.80.Bp, 98.80.Cq, 98.80.Jk}%
\maketitle

\section{Introduction}
The recent discovery of a new scalar like  particle with mass  $125-126\, {\rm GeV}$ in the experiments ATLAS and CMS of the Large Hadron Collider \cite{:2012gk,:2012gu} has profound implications not only in particle physics also in cosmology.
In the particle physics the Higgs field is the key to explain the origin of the mass of all the massive particles known through the Higgs mechanism \cite{Higgs1964,Higgs1964a}.
In the Standard Model the electroweak interactions are described by
a gauge field theory based on the $SU(2)_L\times U(1)_Y$ symmetry
group. 
The spontaneous breaking of the electroweak symmetry through the introduction of a complex scalar doublet field leads to the generation of the $W^\pm$ and $Z$ boson masses.
The Higgs field also gives mass to all fermions through Yukawa interaction \cite{Glashow1961,Weinberg1967}.
However such mechanism of spontaneous symmetry breaking has not been verified experimentally.
The mass of the SM Higgs boson is not predicted by the theory, however the precision electroweak measurements suggest that the mass of the Higgs boson $m_H<152\, {\rm GeV}$ \cite{Barate2003}.
Beyond Standard Model the electroweak symmetry breaking has been studied in the context extra dimensional space-time in \cite{ArkaniHamed:1998rs,Antoniadis1990,Csaki2005}.

In the cosmological context the Higgs particle could play the role of inflaton \cite{Bezrukov:2007ep}. However in \cite{CervantesCota:1995tz,Bezrukov:2007ep} showed that the SM Higgs field cannot act effectively as the inflaton. However this problem can be solved through a non-minimal coupling of the Higgs field with the curvature $R$ \cite{CervantesCota:1995tz,Bezrukov:2007ep}. Other variants of Higgs inflation in 4D theory have been investigated in \cite{Kamada2011,Kamada2012}.
In usual Higgs inflation the value of the coupling constant $\xi$ of the Higgs field must be smaller than $10^{-13}$ for to suppress the amplitude of the curvature perturbation from the inflaton quantum fluctuations  \cite{Salopek1992}. This value of the self-coupling constant implies that the mass of the Higgs particle is $m_H\sim 10^{13}\,{\rm GeV}$ which is close to GUT scale.

In this paper we analyze slow-roll
inflation in Randall-Sundrum type II scenario with a Higgs field confined on the brane.
The Randall-Sundrum braneworld type I was motivated originally as solution to the hierarchy problem between the electroweak  and Planck scales \cite{Randall:1999ee}. The second Randall-Sundrum model was proposed as an alternative mechanism to the Kaluza-Klein compactifications \cite{Randall1999a}.  The Randall-Sundrum type II braneworld   have been intensively studied in the last years, among other reasons, by its appreciable impact in the inflationary scenarios \cite{Hawkins:2000dq,Huey2001,Huey2002}.

Using the covariant formalism \cite{Shiromizu2000}, which relate the four and
five-dimensional space-time, we obtain the effective Einstein equations on the brane:
\be G_{\mu\nu}=-\Lambda_4g_{\mu\nu}+\kappa^2T_{\mu\nu}+\kappa_{(5)}^4S_{\mu\nu}-{\cal E}_{\mu\nu}\label{2.3}
\ee
where $G_{\mu\nu}$ is the usual Einstein tensor, $g_{\mu\nu}$ is the 4-dimensional metric on the brane, $\kappa^2$ is the 4-dimensional gravitational coupling and $\Lambda_4$ is the  effective cosmological constant induced on the brane. The effective Einstein equation on the brane \eqref{2.3} differ of its usual form, the new term $S_{\mu\nu}$ represent quadratic corrections in the matter variables and  ${\cal E}_{\mu\nu}$  is related with projection of the 5-dimensional Weyl's tensor on the brane \cite{Shiromizu2000}.
In a cosmological scenario where the metric induced
on the brane is a spatially flat metric of Friedmann-
Robertson-Walker model, we obtain the following Friedmann equation on the brane (see \cite{Binetruy2000}) 
\begin{equation}\label{Frieeq}
H^2 = \frac{\kappa^2}{3}\rho
\left[ 1 + \frac{\rho}{2\lambda} \right]
    + \frac{\Lambda_4}{3} + \frac{{\cal C}}{a^4} \,,
\end{equation}
where $\kappa^2=8\pi G=8\pi/ M^2$ and $M=1.22\times10^{19}\,{\rm GeV}$
is the four-dimensional Planck mass. The four-dimensional cosmological constant
$\Lambda_4$ induced on the brane is given by
\begin{equation}\label{finetun}
\Lambda_4=\frac{1}{2}\left[\Lambda_5+\kappa^2\lambda\right]
\end{equation}
The Planck masses $M$ and $M_5$ are related through
\begin{equation} 
M=\sqrt{\frac{3}{4\pi\lambda}}M_5 ^3 
\end{equation}
In the braneworld scenarios the Friedmann equation is drastically modified at high
energies. In the regimen of strong brane corrections the $\rho^2$ term in the Friedmann equation \eqref{Frieeq} dominates in the early universe.
The quadratic modification in the Friedmann equation produces increase in the friction on scalar field.
As result in braneworld cosmology the inflation is possible for a wider class of potentials than in
standard cosmology \cite{Hawkins:2000dq}. In other words, this makes slow-roll inflation possible even for potentials are not sufficiently flat from view point of the standard cosmology \cite{Maartens2001,Copeland2001,Huey2001}. The brane corrections to the dynamic of the inflation have been investigated \cite{Maartens2001,Maartens2000,Hawkins:2000dq}.

The high energy modification must occur before nucleosynthesis to recover the observational successes of General Relativity, the cosmological observations impose the lower limit $\lambda\geq (1 {\rm MeV})^4$.
The final term in \eqref{Frieeq} usually called dark radiation term is related with projection of the Weyl's tensor on the brane and it represents the influence of bulk gravitons on the brane see \cite{Maartens:2003tw}. Where ${\cal C}$ is  an integrations constant obtained from analysis of the bulk equations, also this constant can take any sign. In \cite{Bowcock:2000cq} was proved that if
the bulk  is an $AdS_5$, this constant is zero. But if  the bulk is AdS-Schwarzschild, the constant ${\cal C}$ is non-zero and it is possible to relate ${\cal C}$ with the mass of the black hole on the bulk. The dark radiation term in \eqref{Frieeq} is constrained by nucleosynthesis and CMB observations to be no more than $5\%$ of the radiation energy density \cite{Ichiki:2002eh}.
Imposing a fine-tuning between the cosmological constant $\Lambda_5$ of the bulk $AdS_5$ and the positive tension of the brane on the relationship \eqref{finetun}, we can get a cosmological constant on the brane $\Lambda_4=0$.
In the following in this paper,
we will assume $\Lambda_4=0$ and ${\cal C}=0$\footnote{During inflation the dark radiation term is rapidly diluted therefore we can neglect it.}. 
In this case the Friedmann equation \eqref{Frieeq} can be written in the following form
\begin{equation}
\label{Frieeq1}
H^2 = \frac{\kappa^2}{3} \, \rho \, \left[ 1 + \frac{\rho}{2\lambda}\right]
\end{equation}
In the brane scenarios we can obtain inflation on the brane of two different ways.
In the scenarios with a bulk empty  the inflation must be driven by a scalar field trapped on the brane\footnote{In this kind of models the bulk has only a negative cosmological constant.}.
However when the bulk contains a scalar field, it is possible to induce inflation on the brane through the effective projection of the $5D$ scalar field \cite{Himemoto:2000nd,Kobayashi:2000yh}.
In this paper we use the first mechanism to obtain inflation on the brane.  

\section{Higgs Inflation}
\subsection{Higgs field  on the brane}
Here we consider the simple theory of a real scalar field $\Phi$ confined to the brane with Lagrangian
\begin{equation}\label{Lagra}
{\cal L}=\frac{1}{2}g^{\mu\nu}D_\mu \Phi D_\nu \Phi+\xi\left(\Phi^2-\chi^2\right)^2
\end{equation}
Where $g^{\mu\nu}$ is the metric induced on the brane, $\Phi$ is the  Higgs field, the constant $\chi$ is its  vacuum expectation
value and $\xi$ is the self-coupling constant. In the Standard Model  the mass of the Higgs boson is given by $m_H=\sqrt{\frac{\xi}{2}}\chi$ and its vacuum expectation value  $\chi=(\sqrt{2}G_F)^{-1/2}\approx246 {\rm GeV}$ which is fixed by Fermi coupling $G_F$. Like $\xi$ is unknown, the value of the  Higgs boson
mass $m_H$ cannot be predicted.
The  Higgs potential 
\begin{equation}\label{potential}
V(\Phi)=\xi\left(\Phi^2-\chi^2\right)^2
\end{equation}
has been widely studied in the literature as mechanism symmetry breaking \cite{Higgs1964,Higgs1964a,Higgs1966} or cosmological inflation \cite{Bezrukov:2007ep,Salopek1989,Germani2010,Nakayama2010}.
In the braneworld context the Higgs field could be associated with a spontaneous breaking of 4-dimensional symmetry on the brane\footnote{The electroweak symmetry breaking has been explored in the context of the braneworld in \cite{ArkaniHamed:1998rs,Csaki2005}.}.
The scalar field on the brane satisfies the usual Klein-Gordon equation 
\begin{equation}
\label{eqom}
\ddot{\Phi} + 3H \dot{\Phi} +V'(\Phi) = 0 \,.
\end{equation}
For large values of the field $\chi\ll\Phi$ the potential \eqref{potential} can be approximated by
\begin{equation}\label{potential1}
V(\Phi)=\xi\Phi^4
\end{equation}
In this class of potential there is not symmetry breaking.
The inflationary dynamic of the power law potential in braneworld has been investigated in \cite{Nunes2002}. 

The chaotic inflation models $m\Phi^2$ and $\lambda\Phi^4$ in standard cosmology 
have been widely criticized to require super-Planckian field values
to solve the  flatness and anisotropies cosmic microwave background problems.
The problem with super-Planckian field values is that usually we expects non-renormalizable quantum corrections $\sim\left(\Phi/M\right)^a$, $a>4$ to completely dominate the potential, thus we have not control over the potential, destroying the flatness of the potential required for inflation from view point of the standard cosmology. However in \cite{Maartens2000} proved that  chaotic inflation model $m\Phi^2$ in brane framework  can solve these problems for field value below Planck scale. In the next section we proved that an inflation models like $\lambda\Phi^4$ in brane framework do not require super-Planckian field values to solve the  flatness and anisotropies microwave background problems.
 
\subsection{Slow-roll Higgs inflation}
In this section we analyze the slow-roll inflation for a Higgs field confined on the brane.
Analogously to the standard slow-roll inflation based on the General Relativity,
we can define the slow-roll parameter on the brane. The quadratic correction to the Friedmann equation \eqref{Frieeq1} at high energies also modified the standard slow-roll parameter \cite{Maartens2000,Huey2001}.
 
\begin{equation}\label{epsilon}
\epsilon=\frac{1}{2\kappa^2}\left(\frac{V'}{V}\right)^2\left[\frac{4\lambda(1+V)}{(2\lambda+V)^2}\right] 
\end{equation}
\begin{equation}\label{eta}
\eta=\frac{1}{\kappa^2}\frac{V''}{V}\left[\frac{2\lambda}{2\lambda+V}\right]
\end{equation}
The modifications to usual slow-roll parameters are contained in the square brackets of above expressions.
For the potential \eqref{potential1} the slow-roll parameter are given by
\begin{equation}\label{param1}
\epsilon=\frac{32 \lambda  \left(\lambda +\xi  \Phi ^4\right)}{\kappa ^2 \Phi ^2 \left(2 \lambda +\xi  \Phi ^4\right)^2}
\end{equation}
\begin{equation}\label{param2}
\eta=\frac{24 \lambda }{\kappa ^2 \Phi ^2 \left(2 \lambda +\xi  \Phi ^4\right)}
\end{equation}
During slow-roll brane inflation, under approximation $\lambda\ll V(\Phi)$, from \eqref{param1} we have
\begin{equation}
\epsilon \simeq \frac{32 \lambda }{\kappa ^2 \xi  \Phi ^6}
\end{equation}
Inflation ends when $\epsilon = 1$ and this occurs for
\begin{equation}
\Phi_{\rm end}=\sqrt[6]{\frac{32\lambda}{\kappa^2\xi}}
\end{equation}
The potential at the end of inflation
\begin{equation}
\label{Vend}
V_{{\rm end}}=\xi\left(\frac{32\lambda}{\kappa^2\xi}\right)^{2/3}
\end{equation}
The amount of inflation is given by the number $N\approx\int_{t_i} ^{t_f}Hdt$ of e-folds of the scale factor.
The number of $e$-folds also is modified on the brane
\begin{equation}\label{efoldings}
N =- \kappa^2 \int_{\phi_i} ^{\phi_{{\rm end}}} \frac{V}{V'}\left[1 +
\frac{V}{2\lambda}\right] \, d\phi 
\end{equation}
The modifications within square brackets to usual number e-folds allow to obtain more inflation between any two values of the field $\Phi$.   
At very high energies $\lambda\ll V$ we have
\begin{equation}\label{efoldings}
N\approx-\frac{\kappa^2}{2\lambda}\int_{\phi_i} ^{\phi_{{\rm end}}}\frac{V^2}{V'} d\phi 
\end{equation}
For the potential \eqref{potential1} we find
\begin{equation}\label{efold}
N=\frac{2}{3}\left[\left(\frac{V_i}{V_{\rm end}}\right)^{3/2}-1\right]
\end{equation}
The amplitude of the scalar and tensor perturbations produced during
inflation are given \cite{Maartens2000,Huey2001}
\begin{equation}
A_{\rm S}^2 = \frac{\kappa^6}{75\pi^2} \, \frac{V^3}{V'^2} \left(\frac{2\lambda+V}{2\lambda}\right)^3 
\label{scalarpert}
\end{equation}
\begin{equation}
A_{\rm T}^2 = \frac{\kappa^4}{150\pi^2} V \left(\frac{2\lambda+V}{2\lambda}\right)
\label{tensoramp}
\end{equation}
At very high energies the amplitude of the scalar perturbation can be approximated by
\begin{equation}\label{ampscal}
A_{\rm S}^2\approx \frac{\kappa^6}{600\pi^2\lambda^3}\, \frac{V^6}{V'^2}
\end{equation}
and we find that
\begin{equation}
\label{amplscalar}
A_{\rm S}^2\approx \frac{256\xi}{75\pi^2}\left(1+\frac{3}{2}N\right)^3
\end{equation}
This result is independent of the brane tension, then through \eqref{amplscalar} we can determine the coupling constant of the Higgs field. 
\begin{equation}
\xi=\frac{75\pi^2 A_{\rm S}^2 }{256\left(1+\frac{3}{2}N\right)^3}
\end{equation}
From CMB normalization \cite{Bunn:1996da} we know  that amplitude of the scalar perturbation  $A_{\rm S}=2\times10^{-5}$, for $N=55$  we have that $\xi\sim10^{-12}$ which is close to value obtained in \cite{Salopek1992}.

Using the equation \eqref{efoldings} one has
\begin{equation}
N=-\frac{\kappa^2\xi}{48\lambda}\left(\Phi_{\rm end} ^6-\Phi_i ^6\right)
\end{equation}
Hence, it is required that $\Phi_i\sim10^{2}M_5$, to get $N=55$. If the 5-dimensional Planck scale takes values $M_5<10^{17}\, {\rm GeV}$ the inflation occur for  Higgs field value less than the 4-dimensional Planck scale $\Phi_i<10^{19}\,{\rm GeV}$. For example if we assume that $M_5$ is close to the electroweak scale $m_{EW}\sim1\, {\rm TeV}$ we find that the inflation occur  for Higgs field value $\Phi_i<100\,{\rm TeV}$, which is much below the 4-dimensional Planck scale\footnote{The mechanism $M_5\sim m_{EW}$ was suggested like solution to the hierarchy problem between the electroweak and Planck scales \cite{ArkaniHamed:1998rs}.}.
 
The scale-dependence of the perturbations is given by mean of the spectral indices \cite{Maartens2000}.
\begin{equation}
n_S-1=\frac{d \ln A_{\rm S}^2}{d\ln k}\approx-6\epsilon+2\eta
\end{equation}
\begin{equation}
n_T=\frac{d \ln A_{\rm T}^2}{d\ln k}\approx-2\epsilon
\end{equation}
For the potential \eqref{potential1} we find that
\begin{equation}\label{indS}
n_S\approx1-\frac{9}{2+3N}
\end{equation}
\begin{equation}
n_T\approx-\frac{6}{2+3N}
\end{equation}
For $N=55$, it implies that $n_S\approx0.947$ which is close to $n_S=1.17\pm0.31$ from COBE constraints \cite{Gorski1994}. 
In \eqref{indS} we can see that the spectral index tend to
Harrison-Zel'dovich spectrum corresponding to $n_S\rightarrow1$.

In \cite{Ford:1987de} was proved that the density of particles produced after inflation is given by
\begin{equation}
\rho_r ^{\rm end}=0.01g_pH_{\rm end} ^4
\end{equation}
where $\rho_r ^{\rm end}$ and $ H_{\rm end} ^4$ correspond to  values of the energy density  of relativistic particles and Hubble parameter at the end of inflation,  the number of scalar fields involved in particle production take values $10\leq g_p\leq100$.
Using the equations \eqref{Frieeq1} and \eqref{Vend} we find  the ratio between the energy density of relativistic particles and the energy density of the scalar field
\begin{equation}
\frac{\rho_r ^{\rm end}}{\rho_\Phi ^{\rm end}}\approx\frac{10 g_p\xi}{36}
\end{equation}
which depend of the coupling constant of the Higgs field and increase with the number of fields
involved in gravitational production $g_p$. The self-coupling constant computed previously lead to ratio between the energy density of relativistic particles and the energy density of the scalar field $\rho_r ^{\rm end}/\rho_\Phi ^{\rm end}\sim10^{-11}$ when $g_p=100$.

Now let us find the dependence of the Higgs field with the cosmological time.
During slow roll inflation $\dot{\Phi}^2\ll V(\Phi)$  the Friedmann equation can be expressed as a function of the scalar field.	
\begin{equation}\label{Fried3}
H(\Phi)^2=\frac{\kappa^2}{3}V(\Phi)\left(1+\frac{V(\Phi)}{2\lambda}\right)
\end{equation}
At high energies, the above equation can be approximated by
\begin{equation}\label{Fried4}
H(\Phi)=\frac{\kappa}{\sqrt{6\lambda}}V(\Phi)
\end{equation}
such approach is valid only if $\lambda\ll V(\Phi)$.
For simplicity we can write the Friedmann  and Klein-Gordon equations  in the Hamilton-Jacobi form \cite{Lidsey1991b,Salopek1989}.
\begin{equation}
H'a'=-\frac{\kappa^2}{2}H a 
\end{equation}
\begin{equation}
H'=-\frac{\kappa^2}{2}\dot{\Phi}
\end{equation}
The first Hamilton-Jacobi equation can be integrated
\begin{equation}
a(\Phi)=\exp\left[-\frac{\kappa^2}{2}\int\frac{H}{H'}d\Phi\right]
\end{equation}
Substituting the potential \eqref{potential1} into the Friedmann equation \eqref{Fried4}  and integrating in above equation we find
\begin{equation}\label{sol1}
a(\Phi)=C\exp\left[-\frac{\kappa^2}{16}\Phi^2\right]
\end{equation}
Known the scale factor dependence $a(\Phi)$, we can determine the dependence $\Phi(t)$ through the second Hamilton-Jacobi equation which takes the following form  
\begin{equation}\label{Fried5}
\dot{\Phi}-\frac{8\xi}{\kappa\sqrt{6\lambda}}\Phi^3=0
\end{equation}
This equation can be easily integrated 
\begin{equation}\label{sol2}
\Phi(t)=\left(\Phi_0 ^{-2}-\frac{16\xi}{\kappa\sqrt{6\lambda}}(t-t_0)\right)^{-1/2}
\end{equation}
Where $\Phi_0$ is the initial value of the Higgs field corresponding to $\Phi(t_0)$.
Thus we have found the dependence of the Higgs field with the cosmological time.
\section{Disscusion}
In this paper we have analyzed the slow-roll Higgs inflation in brane framework.
We proved that the inflation can be driven by Higgs field with a $\Phi^4$-potential and to produce cosmological perturbations in accordance with observations from COBE. The inflation on the brane occurs for field values below the 4-dimensional Planck scale. This result differ of standard inflation where the potential $\lambda\Phi^4$ require super-Planckian field values.
The compute of the amplitude of the scalar and tensor perturbation allowed to determine the value of the self-coupling constant of the Higgs field $\xi\sim10^{-12}$ which is close to value obtained in \cite{Salopek1992}. This relationship
 between self-coupling constant of the Higgs field and the amplitude of scalar perturbations imply a connection between Higgs particle and the large structures in the universe. The small values of the self-coupling constant of the Higgs field also to suppress the amplitude of the perturbations produced during the inflation also it leads to a very small ratio between the energy density of relativistic particles and the energy density of the scalar field.
Through the Hamilton-Jacobi formalism we established the framework to determine the scale factor dependence with the scalar field and the  scalar field dependence with the cosmological time, to find this dependence allow us to know the behavior of this model.
\bibliography{bib}
\bibliographystyle{apsrev}

\end{document}